\documentclass[11pt]{article}
\usepackage{amsmath,amssymb,amsfonts}

\newtheorem{theorem}{Theorem}
\newtheorem{proposition}{Proposition}

\def\D{{\cal D}}
\def\R{{\mathbb R}}
\def\C{{\mathbb C}}

\def\H{{\mathbb H}}
\def\u{{\bf 1}}
\def\i{{\bf i}}
\def\j{{\bf j}}
\def\k{{\bf k}}
\def\Re{{\mathrm{Re}\, }}

\def\tilde{\widetilde}

\begin{document}

\title{The Moutard transformation for the~Davey--Stewartson II equation
and its geometrical meaning
\thanks{The work was supported by RSCF (grant 19-11-00044).}}
\author{Iskander A. TAIMANOV
\thanks{Novosibirsk State University, 630090 Novosibirsk, and Institute of Mathematics, 630090 Novosibirsk, Russia;
e-mail: taimanov@math.nsc.ru}}
\date{}
\maketitle

\section{Main results}
\label{sec1}

In this article we construct the Moutard type transformation for solutions of the
Davey--Stewartson II equation \cite{DS}
\begin{equation}
\label{ds2}
U_t = i(U_{zz}+U_{\bar{z}\bar{z}} + (V+\bar{V})U), \ \ \ V_{\bar{z}} = 2(|U|^2)_z
\end{equation}
which is the compatibility condition for linear problems
$$
\D \Psi =0 , \ \ \ \partial_t \Psi = A\Psi
$$
where
\begin{equation}
\label{dirac1}
\cal D = \left(
\begin{array}{cc}
0 & \partial \\
-\bar{\partial} & 0
\end{array}
\right) + \left(
\begin{array}{cc}
U & 0 \\
0 & \bar{U}
\end{array}
\right)
\end{equation}
is a two-dimensional Dirac operator with a complex-valued potential $U$ with
$\partial = \frac{1}{2}\big(\frac{\partial}{\partial x} - i\frac{\partial}{\partial y}\big)$,
\begin{equation}
\label{operators1}
A = i\left(\begin{array}{cc} -\partial^2 - V & \bar{U}\bar{\partial} - \bar{U}_{\bar{z}} \\
U\partial - U_z & \bar{\partial}^2 + \bar{V} \end{array}\right),
\end{equation}
and
$$
\Psi =
\left(\begin{array}{cc} \psi_1 & -\bar{\psi}_2 \\
\psi_2 & \bar{\psi}_1 \end{array} \right).
$$
It is also presented as the compatibility condition of the related linear problems
$$
\D^\vee \Phi = 0, \ \ \Phi_t = A^\vee \Phi,
$$
where
\begin{equation}
\label{dirac2}
\D^\vee = \left(\begin{array}{cc} 0 & \partial \\
-\bar{\partial} & 0 \end{array}\right) +
\left(\begin{array}{cc} \bar{U} & 0 \\
0 & U \end{array}\right),
\end{equation}
\begin{equation}
\label{operators2}
A^\vee = - i\left(\begin{array}{cc} -\partial^2 - V & U\bar{\partial} - U_{\bar{z}} \\
\bar{U}\partial - \bar{U}_z & \bar{\partial}^2 + \bar{V} \end{array}\right)
\end{equation}
and
$$
\Phi = \left(\begin{array}{cc} \varphi_1 & -\bar{\varphi}_2 \\
\varphi_2 & \bar{\varphi}_1 \end{array} \right).
$$

It is reasonable to consider $\Psi$ and $\Phi$ as $\H$-valued functions where
$\H$ is the linear space of all quaternions which are realized by matrices of the form
$\left(\begin{array}{cc} a & b \\ -\bar{b} & \bar{a} \end{array}\right)$ with $
a,b \in \C$.

To every pair $\Psi$ and $\Phi$ of $\H$-valued functions
we correspond the $1$-form $\omega$:
$$
\omega(\Phi,\Psi) =
-\frac{i}{2}\left(\Phi^\top \sigma_3 \Psi + \Phi^\top \Psi\right) dz -
\frac{i}{2}\left(\Phi^\top \sigma_3 \Psi - \Phi^\top \Psi\right) d\bar{z},
$$
where $X \to X^\top$ is the transposition of a matrix $X$, and
$\sigma_3= \left(\begin{array}{cc} 1 & 0 \\ 0 & -1 \end{array}\right)$ is the Pauli matrix.
If $\Psi$ and $\Phi$ satisfy the Dirac equations (\ref{dirac1}) and (\ref{dirac2}) then the forms $\omega(\Phi,\Psi)$ and $\omega(\Psi,\Phi)$ are closed and, in particular,
there is defined
an $\H$-valued function
\begin{equation}
\label{sigma}
S(\Phi,\Psi)(z,\bar{z}) = \Gamma \int \omega(\Phi,\Psi).
\end{equation}
Here
$\Gamma = \left(\begin{array}{cc} 0 & 1 \\ -1 & 0 \end{array}\right) = i\sigma_2$,
where $\sigma_2$ is thw Pauli matrix. Formula (\ref{sigma}) gives the spinor (Weierstrass) representation of a surface in the four-space
$\R^4 = \H$ with $z$ a conformal parameter on it.  This surface is defined up to a translation, i.e. up to an integration constant $q \in H$.

Let us define an $\H$-valued function
\begin{equation}
\label{kmatrix}
K(\Phi,\Psi) =  \Psi S^{-1}(\Phi,\Psi)\Gamma \Phi^\top\Gamma^{-1} =
\left(\begin{array}{cc} i\bar{W} & a \\ -\bar{a} & -iW \end{array}\right).
\end{equation}

In \cite{MT} is was shown that

{\sl given $\Psi_0$  and $\Phi_0$, the solutions of (\ref{dirac1}) and (\ref{dirac2}),
for every pair $\Psi$ and $\Phi$  of solutions of the same equations
the $\H$-valued functions
\begin{equation}
\label{moutard1}
\begin{split}
\widetilde{\Psi} =  \Psi - \Psi_0 S^{-1}(\Phi_0,\Psi_0) S(\Phi_0,\Psi), \
\widetilde{\Phi} =  \Phi - \Phi_0 S^{-1}(\Psi_0,\Phi_0) S(\Psi_0,\Phi)
\end{split}
\end{equation}
satisfy the Dirac equations
$$
\widetilde{\D}\widetilde{\Psi} = 0, \ \ \ \ \widetilde{\D}^\vee \widetilde{\Phi} = 0
$$
for the operators $\widetilde{\D}$ and $\widetilde{\D}^\vee$ with the potential
\begin{equation}
\label{newpotential}
\widetilde{U} = U + W,
\end{equation}
where $W$ is defined by the formula (\ref{kmatrix}) for $K(\Phi_0,\Psi_0)$.
Here it is assumed that
\begin{equation}
\label{asym}
\Gamma S^{-1}(\Phi_0,\Psi_0)\Gamma = (S^{-1}(\Psi_0,\Phi_0))^\top
\end{equation}
which is achieved by a convenient choice of integration constant in the definition of $S(\Psi_0,\Phi_0)$ via (\ref{sigma}).
}

This Moutard type transformation of the Dirac operators was derived \cite{MT} and  generalizes the transformation derived in \cite{C} for the case of $U = \bar{U}$ and $\Psi_0 = \Phi_0$. In \cite{C} the Moutard transformation was also
extended to a transformation of solutions of the modified Novikov--Veselov (mNV) equation.

Similarly to the case of the mNV equation the transformation (\ref{moutard1}) and (\ref{newpotential})
is extended to a transformation of solutions of solutions of DSII.
For that let us replace in the definition of $K(\Phi,\Psi)$ the function $S(\Phi,\Psi)$ by
$$
S(\Phi,\Psi)(z,\bar{z},t) = \Gamma \int \omega(\Phi,\Psi) + \Gamma \int \omega_1(\Phi,\Psi),
$$
where
$$
\omega_1(\Phi,\Psi) = 
\left(\left[
\Phi^\top_z \left(\begin{array}{cc} 1 & 0 \\ 0 & 0 \end{array}\right) +
\Phi^\top_{\bar{z}}\left(\begin{array}{cc} 0 & 0 \\ 0 & 1 \end{array}\right)
\right] \Psi
\right.
$$
$$  
\left.
- \Phi^\top\left[
\left(\begin{array}{cc} 1 & 0 \\ 0 & 0 \end{array}\right)\Psi_z +
\left(\begin{array}{cc} 0 & 0 \\ 0 & 1 \end{array}\right)\Psi_{\bar{z}}\right]\right) dt.
$$
We have

\begin{theorem}
If $U$ meets the Davey--Stewartson II equation (\ref{ds2}) and $\Psi$ and $\Phi$ satisfy the equations
$\D \Psi = 0, \Psi_t = A\Psi, \D^\vee \Phi =0, \Phi_t = A^\vee \Phi$, then the Moutard transformation (\ref{newpotential}) of $U$ gives the solution $\tilde{U}$ of the DSII equation
$$
\tilde{U}_t = i(\tilde{U}_{zz}+\tilde{U}_{\bar{z}\bar{z}} + 2(\tilde{V}+\bar{\tilde{V}})\tilde{U}), \ \ \
\tilde{V}_{\bar{z}} = (|\tilde{U}|^2)_z
$$
with
\begin{equation}
\label{newv}
\tilde{V} = V + 2ia_z
\end{equation}
where $a$ is given by (\ref{kmatrix}).
\end{theorem}

The geometrical meaning of this transformation is as follows:

for every fixed $t$ the spinors $\Psi$ and $\Phi$ determine a surface $\Sigma_t$ in $\R^4$ via the Weierstrass (spinor) representation \cite{K1,T1,Trms} and $U$ is the potential of such a representation.
The family of surfaces $\Sigma_t$ evolves via the DSII equation \cite{K2,Taimanov2006}.
To every surface from this family we apply a composition of the inversion centered at the origin and the reflection
$(x_1,x_2,x_3,x_4) \to (-x_1,-x_2,-x_3,x_4)$ and obtain a new surface $\tilde{\Sigma}_t$ for which 
$\tilde{U}$ is the potential of its spinor representation \cite{MT}.  The resulted family of surfaces also evolves via the DSII equation.

Starting with a family of smooth surfaces and the corresponding smooth potentials $U$ we may construct singular up solutions of (\ref{ds2}). Indeed, when $\Sigma_t$ passes through the origin the function $\tilde{U}$ loses continuity or regularity because the origin is mapped into infinity by the inversion.

One of the simplest applications of Theorem 1 consists in constructing exact solutions to (\ref{ds2})
from holomorphic functions. In this case we start from the trivial solution $U=V=0$ for which $\Psi$ and $\Phi$ are defined by holomorphic data and, by Theorem 1, derive nontrivial solutions to (\ref{ds2}). For instance, we have

\begin{theorem}
Let $f(z,t)$ be a function which is holomorphic in $z$ and satisfies the equation
$$
\frac{\partial f}{\partial t} = i\frac{\partial^2 f}{\partial z^2}.
$$
Then
$$
U = \frac{i(zf^\prime - f)}{|z|^2 + |f|^2}, \ \
V = 2ia_z,
$$
where
$$
a = -\frac{i(\bar{z} + f^\prime) \bar{f}}{|z|^2 + |f|^2}
$$
satisfy the Davey--Stewartson II equation (\ref{ds2}).
\end{theorem}

Geometrically we have a deformation of graphs $w=f(z,t)$ which are minimal surfaces in $\R^4 = \C^2$ and whenever $f(z,t)$ vanishes at $z=0$
the graph passes through the origin and the solution $\tilde{U}$ loses continuity or regularity.
Some examples are exposed in \S \ref{examples}.

Let us consider the following change of variables:
$$
X = 2y, \ \ Y = 2x.
$$
In their terms the Davey--Stewartson II equation (\ref{ds2}) takes the form
\begin{equation}
\label{ozeq}
iU_t -U_{XX} + U_{YY} = -4|U|^2U + 8\varphi_X U,
\end{equation}
$$
\Delta\varphi = \frac{\partial^2 \varphi}{\partial X^2}
+ \frac{\partial^2 \varphi}{\partial Y^2} = \frac{\partial}{\partial X} |U|^2,
$$
where $\Re V = 2|U|^2 - 4 \varphi_X$, $\varphi_X = \frac{\partial \varphi}{\partial X}$.
We have to distinguish this version of DSII equation which is called focusing from
the defocusing case for which the right hand side has the opposite sign.

Ozawa constructed a blow-up solution to (\ref{ozeq}) for which the inital data are smooth and fast decaying and in finite time $\|U\|^2$, the squared $l_2$-norm tends to the Dirac distribution \cite{Ozawa}.

Theorem 2 allows construct solutions for which regularity is lost in finite time. For instance, the simplest one of them has the form
$$
U = \frac{i(z^2-2it+iT)}{|z|^2 + |z^2 + 2it-iT|^2}.
$$
It is regular for $t \neq T/2$ and $U \sim i e^{2i \phi}$ for $t=T/2$, where
$z = r e^{i\phi}$.

We discuss in more details these solutions in \S\S \ref{examples} and \ref{remarks}.

This article may be considered ass a continuation of \cite{T151,T152} where similar ideas were applied to the modified Novikov--Veselov (mNV) equation. In \cite{T151} we found in terms of the Mobius geometry of $\R^3$ a geometrical interpretation of the Moutard transformation from \cite{C} and in \cite{T152} used it for constructing solutions to mNV which loses regularity in finite time. In \cite{MT} and this article we successively realize such a program for the case of DSII equation.

\section{Surfaces in $\R^4$ via spinors and their soliton deformations}

\subsection{The Weierstrass (spinor) representation of surfaces in the four-space}

For solutions $\psi$ and $\varphi$ to the Dirac equations
$\D\psi=\D^\vee\varphi=0$ the formulas
\begin{equation}
\label{wr2}
x^k = x^k(0) + \int \eta_k, \ \ \ k=1,2,3,4,
\end{equation}
where
$$
\eta_k = f_k dz + \bar{f}_k d\bar{z}, \ \ \ k=1,2,3,4,
$$
$$
f_1 = \frac{i}{2} (\bar{\varphi}_2\bar{\psi}_2 + \varphi_1 \psi_1),
\ \ \ \
f_2 = \frac{1}{2} (\bar{\varphi}_2\bar{\psi}_2 - \varphi_1 \psi_1),
$$
$$
f_3 = \frac{1}{2} (\bar{\varphi}_2 \psi_1 + \varphi_1 \bar{\psi}_2),
\ \ \ \
f_4 = \frac{i}{2} (\bar{\varphi}_2 \psi_1 - \varphi_1 \bar{\psi}_2)
$$
determine an immersed surface in $\R^4$ \cite{K2}.
The induced metric takes the form
$$
e^{2\alpha} dzd\bar{z} =
(|\psi_1|^2+|\psi_2|^2)(|\varphi_1|^2+|\varphi_2|^2)dz d\bar{z}
$$
and the potential $U$ of the Dirac operators relates to the mean curvature vector
${\bf H}$ as follows
$$
|U| = \frac{|{\bf H}| e^\alpha}{2}, \ \ {\bf H} = 2 e^{-2\alpha}\,x_{z\bar{z}},
\ \ x = (x^1,x^2,x^3,x^4) \in \R^4.
$$
The integrals in (\ref{wr2}) are defined up to constants of integration which means that the surface is defined up to translations.

For $\psi=\varphi$ and $U$ real-valued we have $x^4 = \mathrm{const}$ and these formulas give a surface in $\R^3$
\cite{K1}.

In fact these formulas are general and give a representation of any surface with a fixed conformal parameter $z$. For topologically nontrivial surfaces the $\Psi$ and
$\Phi$ functions are sections of spinor bundles and the integral of $|U|^2$ over the surface is just the Willmore functional up to a constant multiple \cite{T1,Taimanov2006} (see also the survey \cite{Trms}).

 For surfaces in $\R^3$ the spinor $\psi$ is uniquely defined  up to a multiplication by $\pm 1$, however for surfaces in $\R^4$ the representations of the same surface are related by gauge transformations of the form $(\psi_1,\psi_2) \to (e^f\psi_1,e^{\bar{f}}\psi_2)$ and
$ (\varphi_1,\varphi_2) \to (e^{-f}\varphi_1,e^{-\bar{f}}\varphi_2)$ where $f$ is a holomorphic function.

\subsection{The geometrical meaning of the Moutard transformation}

Following \cite{T151,MT}, let us identify the four-space with the quaternion space $\H$ and
rewrite the spinor representation in the form
\begin{equation}
\label{spin}
S(\Phi,\Psi) =  \int\left[
i\left(\begin{array}{cc} \psi_1\bar{\varphi}_2 & -\bar{\psi}_2\bar{\varphi}_2 \\
\psi_1 \varphi_1 & -\bar{\psi}_2\varphi_1 \end{array} \right) dz +
i\left(\begin{array}{cc} \psi_2\bar{\varphi}_1 & \bar{\psi}_1\bar{\varphi}_1 \\
- \psi_2 \varphi_2 & -\bar{\psi}_1\varphi_2 \end{array} \right) d\bar{z}\right]
 =
\end{equation}
$$
\int d \left(\begin{array}{cc} ix^3 + x^4 & -x^1-ix^2 \\
x^1-ix^2 & -ix^3+x^4 \end{array}\right)
$$
where the spinors  $\Psi$ and $\Phi$ satisfy the Dirac equations (\ref{dirac1}) and (\ref{dirac2}). It is easy to notice that $S(\Phi,\Psi)$ is the same as in (\ref{sigma}).

The meaning of the Moutard transformation (\ref{moutard1}) and (\ref{newpotential}) is as follows.
We consider $S$ as a surface in $S^4 = \R^4 \cup \{\infty\}$ determined by $\Psi_0$ and $\Phi_0$. Then we apply to $S$
the conformal transformation
$$
S \to \widetilde{S} = S^{-1}.
$$
The resulted surface is determined by the spinors
$$
\widetilde{\Psi}_0 =  \Psi_0 S^{-1}(\Phi_0,\Psi_0), \ \
\widetilde{\Phi}_0 = \Phi_0 S^{-1}(\Psi_0,\Phi_0)
$$
which satisfy the Dirac equations with the potential $\widetilde{U}$.
Since $S$ is defined up to integrations constants we assume here that
$S(\Phi_0,\Psi_0)$ and $S(\Psi_0,\Phi_0)$ are related by (\ref{asym}).

For the real-valued potential $U$ which corresponds to surfaces in $\R^3$ the geometry of the
Moutard transform was  descibed in \cite{T151} and later generalized for the case of
general potentials $U$ in \cite{MT}.

It is worth to explain all operations of $S$ via quaternions.
Let us introduce the following basis in $\H$:
$$
\u = \left(\begin{array}{cc} 1 & 0 \\ 0 & 1 \end{array}\right), \
\i = \left(\begin{array}{cc} 0 & -1 \\ 1 & 0 \end{array}\right), \
\j = \left(\begin{array}{cc} 0 & -i \\ -i & 0 \end{array}\right), \
\k =\left(\begin{array}{cc} i & 0 \\ 0 & -i \end{array}\right).
$$
The surface $S$ of the form (\ref{spin}) as any other matrix from $\H$ is written in coordinates as
$$
S(\Phi,\Psi) = x^1\i + x^2 \j + x^3 \k + x^4\u
$$
and the standard operations are written as follows:
$$
S^{-1} = \frac{1}{|x|^2}(-x^1\i - x^2 \j - x^3 \k + x^4\u), \ \
S^\top = -x^1\i + x^2 \j + x^3 \k + x^4\u.
$$
It is clear from here that in terms of surfaces the transformation $S \to S^{-1}$ is a composition of
the inversion $x \to \frac{x}{|x|^2}$ and the reflection $(x_1,x_2,x_3,x_4) \to (-x_1,-x_2,-x_3,x_4)$.

We remark that $\Gamma = -\i$ and if the functions $S(\Psi,\Phi)$ and $S(\Phi,\Psi)$
defined up to constants are normalized such that they both equal to zero at the same point, then
$$
S(\Psi,\Phi) = x^1\i + x^2 \j + x^3 \k - x^4\u
$$
which implies (\ref{asym}).

\subsection{The soliton deformation of surfaces via the DSII equation}

The Dirac operators $\D$ and $\D^\vee$ are coming into
the $L,A,B$-triples for the Davey--Stewartson equations.
This means that there are such differential operators $A,B$ and $A^\vee,B^\vee$
that the compatibility conditions for the systems
$$
\D \psi = 0, \ \ \psi_t = A\psi, \ \ \psi = \left(\begin{array}{c} \psi_1 \\ \psi_2 \end{array}\right)
$$
and
$$
\D^\vee \varphi = 0,  \ \ \varphi_t = A^\vee \varphi,
\ \ \varphi = \left(\begin{array}{c} \varphi_1 \\ \varphi_2 \end{array}\right)
$$
takes the form of a scalar equation from the Davey--Stewartson hierarchy.
This equation is presented by the triples
$$
\D_t + [\D,A] - B\D = 0, \ \ \ \D^\vee_t + [\D^\vee,A^\vee] - B^\vee\D^\vee = 0.
$$
Hence if $\psi$ and $\varphi$ evolve due to this equation, then by the spinor representation they
determine a soliton deformation $S(\Psi,\Phi,t)$ of the corresponding surfaces.

Such deformations were introduced in \cite{K1}  for surfaces in the three-space  and they correspond to the modified Novikov--Veselov hierarchy, and then in \cite{K2} for surfaces in $\R^4$. In the latter case
the corresponding integrable systems are equations from the Davey--Stewartson hierarchy. Since these equations contain nonlocal terms, it needs to resolve the constraints carefully for having the Willmore functional as the first integral of the system \cite{Taimanov2006}:
$$
\frac{d}{dt} \int |U|^2 \, dx \wedge dy = 0.
$$
For this reason we consider only the DSII equation which corresponds to the $A$-operators (\ref{operators1}) and (\ref{operators2})  and takes the form (\ref{ds2}).

\section{The Moutard transformation for the DSII equation: proof of Theorem 1}

The DSII deformation of surfaces is defined via the deformation of the determining spinors which give us not a surface but its Gauss map. Hence it is defined up to translations. Let us assume that translations are smooth functions of the temporary variable. Then the deformation is written in the form
$$
S(\Phi,\Psi)(Z,\bar{Z},t) = \int_{(P,0)}^{(Z,\bar{Z},t)} (M dz + N d\bar{z} + Qdt),
$$
where $z$ is a conformal parameter in a (simply-connected) domain $\cal{U} \in \C$, $P \in \cal{U}$ is a certain fixed point, and $(Z,\bar{Z}) \in \cal{U}$.

By definiton of the DSII deformation, we have
$$
M =
i\left(\begin{array}{cc} \psi_1\bar{\varphi}_2 & -\bar{\psi}_2\bar{\varphi}_2 \\
\psi_1 \varphi_1 & -\bar{\psi}_2\varphi_1 \end{array} \right), \ \
N =
i\left(\begin{array}{cc} \psi_2\bar{\varphi}_1 & \bar{\psi}_1\bar{\varphi}_1 \\
- \psi_2 \varphi_2 & -\bar{\psi}_1\varphi_2 \end{array} \right).
$$
Since the one-form $Mdz + Nd\bar{z} + Qdt$ has to be closed, we have
$$
\frac{\partial M}{\partial \bar{z}} = \frac{\partial N}{\partial z}, \ \
\frac{\partial M}{\partial t} = \frac{\partial Q}{\partial z}, \ \
\frac{\partial N}{\partial t} = \frac{\partial Q}{\partial \bar{z}}.
$$
The first equation follows from (\ref{dirac1}) and (\ref{dirac2}), and from two other equations we can  determine $Q$ up to functions of $t$.

\begin{proposition}
Let spinors $\Psi$ and $\Phi$ satisfy the Dirac equations $\D\Psi = \D^\vee\Phi =0$ and the evolution equations $\Psi_t = A\Psi$ and $\Phi_t = A^\vee \Phi$, where
$U$ satisfy the DSII equation (\ref{ds2}). Then
the DSII deformation of the surface defined by the spinors $\Phi$ and $\Psi$ takes the form
\begin{equation}
\label{gsigma}
S(\Phi,\Psi) =
\end{equation}
$$
= \int \left[
i\left(\begin{array}{cc} \psi_1\bar{\varphi}_2 & -\bar{\psi}_2\bar{\varphi}_2 \\
\psi_1 \varphi_1 & -\bar{\psi}_2\varphi_1 \end{array} \right) dz +
i\left(\begin{array}{cc} \psi_2\bar{\varphi}_1 & \bar{\psi}_1\bar{\varphi}_1 \\
- \psi_2 \varphi_2 & -\bar{\psi}_1\varphi_2 \end{array} \right) d\bar{z} + Q_0(t)dt + \right.
$$
$$
\left.
\left(\begin{array}{cc}
\psi_{1z} \bar{\varphi}_2 - \psi_1\bar{\varphi}_{2z} +
\psi_2\bar{\varphi}_{1\bar{z}} - \psi_{2\bar{z}}\bar{\varphi}_1
&
\bar{\psi}_1\bar{\varphi}_{1\bar{z}} - \bar{\psi}_{1\bar{z}}\bar{\varphi}_1 +
\bar{\psi}_2\bar{\varphi}_{2z} - \bar{\psi}_{2z}\bar{\varphi}_2
\\
\psi_{1z}\varphi_1 - \psi_1\varphi_{1z} +
\psi_{2\bar{z}}\varphi_2 - \psi_2 \varphi_{2\bar{z}}
&
\bar{\psi}_2\varphi_{1z} - \bar{\psi}_{2z} \varphi_1 +
\bar{\psi}_{1\bar{z}}\varphi_2 - \bar{\psi}_1 \varphi_{2\bar{z}}
\end{array}\right)
dt \right],
$$
where $Q_0(t)$ is a smooth $\H$-valued function of $t$.
\end{proposition}

{\sl Proof.} Since $M_{11} = \bar{N}_{22}, M_{22} = \bar{N}_{11}, M_{12} = -\bar{N}_{21}$, $M_{21} = \bar{N}_{12}$, and $Q$ is a $\H$-valued function, it is enough to consider the equations
$$
\frac{\partial M}{\partial t} = \frac{\partial Q}{\partial z}.
$$

Let us consider $M_{11} = i\psi_1 \bar{\varphi}_2$. By (\ref{operators1}) and (\ref{operators2}),
$$
(\psi_1 \bar{\varphi}_2)_t =
i[(-\psi_{1zz} - V\psi_1) + (\bar{U}\psi_{2\bar{z}} -
\bar{U}_{\bar{z}}\psi_2)]\bar{\varphi}_2 +
i\psi_1[(U\bar{\varphi}_{1\bar{z}}-U_{\bar{z}}\bar{\varphi}_1) +
$$
$$
+ (\bar{\varphi}_{2zz} + V\bar{\varphi}_2)] =
i[(\psi_1\bar{\varphi}_{2zz} -
\psi_{1zz}\bar{\varphi}_2) + (\bar{U}\psi_{2\bar{z}}\bar{\varphi}_2 - \bar{U}_{\bar{z}}\psi_2 \bar{\varphi}_2) +
$$
$$
+ (U\psi_1\bar{\varphi}_{1\bar{z}} - U_{\bar{z}}\psi_1\bar{\varphi}_1)]
$$
We split the right-hand side into three parts and consider each of them separately.
Clearly
$$
\psi_1\bar{\varphi}_{2zz} - \psi_{1zz}\bar{\varphi}_2 =
(\psi_1\bar{\varphi}_{2z} - \psi_{1z}\bar{\varphi}_2)_z.
$$
Let us rewrite (\ref{dirac1}) and (\ref{dirac2}) as
\begin{equation}
\label{diraceq1}
U\psi_1 = -\psi_{2z}, \ \ \bar{U}\psi_2 = \psi_{1\bar{z}}, \ \
U\varphi_2 = \varphi_{1\bar{z}}, \ \ \bar{U} \varphi_1 = -\varphi_{2z}
\end{equation}
and consider two other parts.
By (\ref{diraceq1}) we conclude that
$$
\bar{U}\psi_{2\bar{z}}\bar{\varphi}_2 - \bar{U}_{\bar{z}}\psi_2 \bar{\varphi}_2 =
\bar{U}\psi_{2\bar{z}}\bar{\varphi}_2 -(\bar{U}\psi_2\bar{\varphi}_2)_{\bar{z}}
+ \bar{U}(\psi_2 \bar{\varphi}_2)_{\bar{z}}=
$$
$$
2\bar{U}\psi_{2\bar{z}}\bar{\varphi}_2 + \bar{U}\psi_2\bar{\varphi}_{2\bar{z}} - (\psi_2 \bar{\varphi}_{1z})_{\bar{z}} = \psi_{2\bar{z}}\bar{\varphi}_{1z} + \psi_{1\bar{z}} \bar{\varphi}_{2\bar{z}} - \psi_2\bar{\varphi}_{1z\bar{z}}.
$$
Analogously we infer that
$$
U\psi_1\bar{\varphi}_{1\bar{z}} - U_{\bar{z}}\psi_1\bar{\varphi}_1 =
U\psi_1\bar{\varphi}_{1\bar{z}} - (U\psi_1\bar{\varphi}_1)_{\bar{z}} +
U(\psi_1\bar{\varphi}_1)_{\bar{z}} =
$$
$$
2U \psi_1 \bar{\varphi}_{1\bar{z}} + U\psi_{1\bar{z}}\bar{\varphi} + (\psi_{2z}\bar{\varphi}_1)_{\bar{z}} =
-\psi_{2z}\bar{\varphi}_{1\bar{z}} - \psi_{1\bar{z}}\bar{\varphi}_{2\bar{z}} +
\psi_{2z\bar{z}}\bar{\varphi}_1.
$$
Combining the results of calculations for all three parts we derive
$$
(\psi_1 \bar{\varphi}_2)_t = i(\psi_1\bar{\varphi}_{2z} - \psi_{1z}\bar{\varphi}_2 +
\psi_{2\bar{z}}\bar{\varphi}_1 - \psi_2\bar{\varphi}_{1\bar{z}})_z.
$$
Therefore
$Q_{11} = - (\psi_1\bar{\varphi}_{2z} - \psi_{1z}\bar{\varphi}_2 +
\psi_{2\bar{z}}\bar{\varphi}_1 - \psi_2\bar{\varphi}_{1\bar{z}})+ Q_{0,11}(\bar{z},t)$.
From the equation $\frac{\partial N}{\partial t} = \frac{\partial Q}{\partial \bar{z}}$ it follows that $Q_{0,11}= Q_{0,11}(t)$.
Hence we established (\ref{gsigma}) for $M_{11}$. The formulas for a
all other entries of $M$ and $N$ are derived similarly and we skip these  calculations. Proposition is proved.

Let us consider the $S_t$ term in detail:
$$
\left(\begin{array}{cc}
\psi_{1z} \bar{\varphi}_2 - \psi_1\bar{\varphi}_{2z} +
\psi_2\bar{\varphi}_{1\bar{z}} - \psi_{2\bar{z}}\bar{\varphi}_1
&
\bar{\psi}_1\bar{\varphi}_{1\bar{z}} - \bar{\psi}_{1\bar{z}}\bar{\varphi}_1 +
\bar{\psi}_2\bar{\varphi}_{2z} - \bar{\psi}_{2z}\bar{\varphi}_2
\\
\psi_{1z}\varphi_1 - \psi_1\varphi_{1z} +
\psi_{2\bar{z}}\varphi_2 - \psi_2 \varphi_{2\bar{z}}
&
\bar{\psi}_2\varphi_{1z} - \bar{\psi}_{2z} \varphi_1 +
\bar{\psi}_{1\bar{z}}\varphi_2 - \bar{\psi}_1 \varphi_{2\bar{z}}
\end{array}\right) =
$$
$$
=
\left(\begin{array}{cc} -\bar{\varphi}_{2z} & \bar{\varphi}_{1\bar{z}} \\
-\varphi_{1z} & -\varphi_{2\bar{z}} \end{array}\right)
\left(\begin{array}{cc} \psi_1 & -\bar{\psi}_2 \\
\psi_2 & \bar{\psi}_1 \end{array}\right) +
\left(\begin{array}{cc} \bar{\varphi}_2 & -\bar{\varphi}_1 \\
\varphi_1 & \varphi_2 \end{array}\right)
\left(\begin{array}{cc} \psi_{1z} & -\bar{\psi}_{2z} \\
\psi_{2\bar{z}} & \bar{\psi}_{1\bar{z}} \end{array}\right)
=
$$
\begin{equation}
\label{st2}
=
\left(\begin{array}{cc} 0 & 1 \\ -1 & 0 \end{array}\right)\left[
\Phi^\top_z \left(\begin{array}{cc} 1 & 0 \\ 0 & 0 \end{array}\right) +
\Phi^\top_{\bar{z}}\left(\begin{array}{cc} 0 & 0 \\ 0 & 1 \end{array}\right)
\right] \Psi  +
\end{equation}
$$
+
\left(\begin{array}{cc} 0 & -1 \\ 1 & 0 \end{array}\right)\Phi^\top\left[
\left(\begin{array}{cc} 1 & 0 \\ 0 & 0 \end{array}\right)\Psi_z +
\left(\begin{array}{cc} 0 & 0 \\ 0 & 1 \end{array}\right)\Psi_{\bar{z}}\right].
$$
Analogously we obtain that
$$
S_z = -i \Gamma \Phi^\top \left(\begin{array}{cc} 1 & 0 \\ 0 & 0 \end{array}\right)\Psi, \ \
S_{\bar{z}} = - i \Gamma \Phi^\top \left(\begin{array}{cc} 0 & 0 \\ 0 & -1 \end{array}\right)\Psi.
$$

Let us consider the surface
$$
\tilde{S} = S^{-1}(\Phi,\Psi).
$$
By  \cite[Proposition 1]{MT}, it is obtained from $S = S(\Phi,\Psi)$ as a composition of the inversion and the reflection
$(x^1,x^2,x^3,x^4) \to (-x^1,-x^2,-x^3,x^4)$ and its spinor representation is given by
$$
\tilde{\Phi} = \Phi S^{-1}(\Psi,\Phi), \ \ \tilde{\Psi} = \Psi S^{-1}(\Phi,\Psi) = \Psi S^{-1}.
$$
By (\ref{asym}), we rewrite  the transformation of $\Phi$ as
$$
\tilde{\Phi}^\top = \Gamma S^{-1} \Gamma \Phi^\top.
$$
Let us substitute these formulas for $\tilde{\Phi}$ and $\tilde{\Psi}$ into (\ref{st2}) and obtain the formula for $\tilde{S}_t$ modulo a function $\tilde{Q}_0(t)$.
If we compare it with $S^{-1}_t = -S^{-1}S_tS^{-1}$ we derive by straightforward calculations that these two expressions coincide if and only if
$\tilde{Q}_0(t) = -S^{-1}(z,\bar{z},t)Q_0(t)S^{-1}(z,\bar{z},t)$. For a generic surface there are no such nontrivial functions $Q_0$ and $\widetilde{Q}_0$. Therefore
we have

\begin{proposition}
Let us assume that the conditions of Proposition 1 holds. Given $Q_0=0$, we have
$$
\tilde{S}_t =
$$
$$
\Gamma \left[
\tilde{\Phi}^\top_z \left(\begin{array}{cc} 1 & 0 \\ 0 & 0 \end{array}\right) +
\tilde{\Phi}^\top_{\bar{z}}\left(\begin{array}{cc} 0 & 0 \\ 0 & 1 \end{array}\right)
\right] \tilde{\Psi}  -
\Gamma\tilde{\Phi}^\top\left[
\left(\begin{array}{cc} 1 & 0 \\ 0 & 0 \end{array}\right)\tilde{\Psi}_z +
\left(\begin{array}{cc} 0 & 0 \\ 0 & 1 \end{array}\right)\tilde{\Psi}_{\bar{z}}\right].
$$
and the surface $\tilde{S} = S^{-1}$ of the form (\ref{gsigma}) also deforms via the DSII equation.
\end{proposition}

By \cite{MT}, $\tilde{\Psi}$ and $\tilde{\Phi}$ satisfy the Dirac equations with potential $\tilde{U} = U+W$, where $W$ is given by (\ref{kmatrix}) which we rewrite as
$$
K = \tilde{\Psi}\Gamma \Phi^\top \Gamma^{-1} = \left(\begin{array}{cc} i \bar{W} & a \\ -\bar{a} & -iW \end{array}\right).
$$
We are left to find operators $\tilde{A}$ and $\tilde{A}^\vee$ of the forms (\ref{operators1}) and
(\ref{operators2})  such that
$$
\tilde{\Psi}_t = \tilde{A}\tilde{\Psi}, \ \ \ \tilde{\Phi}_t = \tilde{A}^\vee \tilde{\Phi}.
$$
Since $\tilde{U}$ is known, we have to find $\tilde{V}$. For that let us write down
$$
\tilde{\Psi}_t = (\Psi \tilde{S})_t = \Psi_t \tilde{S} + \Psi \tilde{S}_t = A\Psi \cdot \tilde{S} + \Psi \tilde{S}_t =
$$
$$
= A\tilde{\Psi} - \left[ -i \left(\begin{array}{cc} 1 & 0 \\ 0 & 0 \end{array}\right)(2 \Psi_z \tilde{S}_z  +\Psi \tilde{S}_{zz}) + i \left(\begin{array}{cc} 0 & 0 \\ 0 & 1 \end{array}\right)(2 \Psi_{\bar{z}} \tilde{S}_{\bar{z}}  +\Psi \tilde{S}_{\bar{z}\bar{z}})   + \right.
$$
$$
\left.
+ \bar{U} \left(\begin{array}{cc} 0 & 1 \\ 0 & 0 \end{array}\right)\Psi\tilde{S}_{\bar{z}} +
U \left(\begin{array}{cc} 0 & 0 \\ 1 & 0 \end{array}\right)\Psi\tilde{S}_{z}\right] +
$$
$$
+
\Psi\Gamma \left[
\tilde{\Phi}^\top_z \left(\begin{array}{cc} 1 & 0 \\ 0 & 0 \end{array}\right) +
\tilde{\Phi}^\top_{\bar{z}}\left(\begin{array}{cc} 0 & 0 \\ 0 & 1 \end{array}\right)
\right] \tilde{\Psi}  -
$$
$$
\Psi\Gamma \tilde{\Phi}^\top\left[
\left(\begin{array}{cc} 1 & 0 \\ 0 & 0 \end{array}\right)\tilde{\Psi}_z +
\left(\begin{array}{cc} 0 & 0 \\ 0 & 1 \end{array}\right)\tilde{\Psi}_{\bar{z}}\right].
$$
We substitute into the formula the expressions for derivatives of $\tilde{S}$ and obtain that
$$
\tilde{\Psi}_t = A\tilde{\Psi} + \Lambda \tilde{\Psi} + \Theta
$$
where $\Lambda$ is a matrix which we consider below and
$$
\Theta =
\left(\begin{array}{cc} 0 & 0 \\ 0 & -1 \end{array}\right)\Psi\Gamma\tilde{\Phi}^\top\left(\begin{array}{cc} 1 & 0 \\ 0 & 0 \end{array}\right)\tilde{\Psi}_z +
\left(\begin{array}{cc} -1 & 0 \\ 0 & 0 \end{array}\right)\Psi\Gamma\tilde{\Phi}^\top\left(\begin{array}{cc} 1 & 0 \\ 0 & 0 \end{array}\right)\tilde{\Psi}_{\bar{z}} =
$$
$$
\left(\begin{array}{cc} 0 & 0 \\ 0 & 1 \end{array}\right)K\Gamma\left(\begin{array}{cc} 1 & 0 \\ 0 & 0 \end{array}\right)\tilde{\Psi}_z +
\left(\begin{array}{cc} 1 & 0 \\ 0 & 0 \end{array}\right)K\Gamma\left(\begin{array}{cc} 1 & 0 \\ 0 & 0 \end{array}\right)\tilde{\Psi}_{\bar{z}} =
$$
$$
=
\left(\begin{array}{cc} 0 & k_{11} \bar{\partial} \\ -k_{22} \partial & 0
\end{array}\right)\tilde{\Psi} = i \left(\begin{array}{cc} 0 & \bar{W} \bar{\partial} \\ W \partial & 0
\end{array}\right)\tilde{\Psi}.
$$
The matrix $\Lambda$ is presented by the following sum
$$
\Lambda = \Psi\Gamma \left[
\tilde{\Phi}^\top_z \left(\begin{array}{cc} 1 & 0 \\ 0 & 0 \end{array}\right) +
\tilde{\Phi}^\top_{\bar{z}}\left(\begin{array}{cc} 0 & 0 \\ 0 & 1 \end{array}\right)
\right] + 2\left(\begin{array}{cc} 1 & 0 \\ 0 & 0 \end{array}\right)\Psi_z\Gamma\tilde{\Phi}^\top
\left(\begin{array}{cc} 1 & 0 \\ 0 & 0 \end{array}\right) +
$$
$$
+
2\left(\begin{array}{cc} 0 & 0 \\ 0 & 1 \end{array}\right)\Psi_{\bar{z}}\Gamma\tilde{\Phi}^\top
\left(\begin{array}{cc} 0 & 0 \\ 0 & 1 \end{array}\right) +
\left(\begin{array}{cc} 1 & 0 \\ 0 & 0 \end{array}\right)\Psi\Gamma\tilde{\Phi}^\top_z
\left(\begin{array}{cc} 1 & 0 \\ 0 & 0 \end{array}\right) +
$$
\begin{equation}
\label{lamb}
+ \left(\begin{array}{cc} 0 & 0 \\ 0 & 1 \end{array}\right)\Psi\Gamma\tilde{\Phi}^\top_{\bar{z}}
\left(\begin{array}{cc} 0 & 0 \\ 0 & 1 \end{array}\right) +
\end{equation}
$$
+ i\bar{U} \left(\begin{array}{cc} 0 & 1  \\ 0 & 0 \end{array}\right)\Psi\Gamma\tilde{\Phi}^\top
\left(\begin{array}{cc} 0 & 0 \\ 0 & -1 \end{array}\right) +
 iU \left(\begin{array}{cc} 0 & 0  \\ 1 & 0 \end{array}\right)\Psi\Gamma\tilde{\Phi}^\top
\left(\begin{array}{cc} 1 & 0 \\ 0 & 0 \end{array}\right).
$$
We conclude that
$$
\tilde{\Psi}_t = i\left(\begin{array}{cc} -\partial^2 - V & \bar{\tilde{U}}\bar{\partial} - \bar{U}_{\bar{z}} \\
\tilde{U}\partial - U_z & \bar{\partial}^2 + \bar{V} \end{array}\right)\tilde{\Psi} + \Lambda\tilde{\Psi}
$$
and hence
$$
\tilde{V} = V + i \Lambda_{11}.
$$
It is easy to notice that not all components of (\ref{lamb}) contributes to
$\Lambda_{11}$ which is equal to
$M_{11}$ where
$$
M = \Psi \Gamma \tilde{\Phi}_z^\top + 2 \left(\begin{array}{cc} 1 & 0 \\ 0 & 0 \end{array}\right)\Psi_z \Gamma \tilde{\Phi}^\top +  \left(\begin{array}{cc} 1 & 0 \\ 0 & 0 \end{array}\right) \Psi \Gamma \tilde{\Phi}^\top_z =
$$
$$
=
2 \left(\begin{array}{cc} 1 & 0 \\ 0 & 0 \end{array}\right)(\Psi \Gamma \tilde{\Phi}^\top)_z +  \left(\begin{array}{cc} 0 & 0 \\ 0 & 1 \end{array}\right) \Psi \Gamma \tilde{\Phi}^\top_z =
$$
$$
=
-2 \left(\begin{array}{cc} 1 & 0 \\ 0 & 0 \end{array}\right)(K\Gamma)_z +  \left(\begin{array}{cc} 0 & 0 \\ 0 & 1 \end{array}\right) \Psi \Gamma \tilde{\Phi}^\top_z.
$$
From the form of $K$ we derive that
$\Lambda_{11} = M_{11} = 2 a_z$ and thereby obtain (\ref{newv}):
$$
\tilde{V} = V + 2ia_z.
$$
This finishes the proof of Theorem 1.

\section{Minimal surfaces in the four-space and exact solutions to the DSII equation}

\subsection{Proof of Theorem 2}

Minimal surfaces  are defined as surfaces with  vanishing mean curvature $H=0$ and in our these are exactly the surfaces which are
determined by (\ref{spin}) where the functions $\psi_1, \bar{\psi}_2, \varphi_1$, and $\bar{\varphi}_2$ are holomorphic.
The holomorphicity conditions mean that $\Psi$ and $\Phi$ satisfy the Dirac equations $\D \Psi = 0$ and $\D^\vee \Phi$ for $U=0$,

Let us consider such holomorphic functions as the initial data (for $t=0$) for solutions of the equations
$$
\psi_{1t} = - i \partial^2 \psi_1, \ \ \bar{\psi}_{2t} = - \partial \bar{\psi}_{2}, \ \
\varphi_{1t} = i \partial^2 \varphi_1, \ \ \bar{\varphi}_{2t} = i \partial^2 \bar{\varphi}_2.
$$
These are exactly the equations $\Psi_t = A \Psi$ and $\Phi_t = A^\vee \Phi$ for $U=V=0$.
The solutions $\Psi(z,\bar{z},t$ and $\Phi(z,\bar{z},t$ to these equations
define a one-parameter family of minimal surfaces $S(z,\bar{z},t)$.
The inversion $S^{-1}$ of these surfaces gives us the Moutard transformation (Theorem 1)  of a trivial solution $U=V=0$ to the DSII equation.

Therefore we have a simple method to write down a family of non-trivial solutions to the  DSII equations which depend on four functional parameters which are holomorphic functions.

We skip the general formulas and demonstrate this method by considering a one-parameter family of solutions.

Let is identify $\R^4$ with $\C^2$ with coordinates $z = x_1 + i x_2, w = x_3 + i x_4$.
For every holomorphic function $f$ its graph of $f$:
$$
w = f(z),
$$
determines a minimal surface in $\R^4$.
The spinor representation (\ref{spin}) of this surface is defined by spinors
$$
\Psi = \left(\begin{array}{cc} 0 & -1 \\ 1 & 0 \end{array}\right), \ \
\Phi = \left(\begin{array}{cc} f^\prime & i \\ i & \bar{f}^\prime \end{array}\right)
$$
and takes the form
$$
S = \left(\begin{array}{cc} i\bar{f} & -z \\ \bar{z} & -if \end{array}\right).
$$
Assume that $f$ also depends on $t$ and satisfies the equation
$$
f_t = i \partial^2 f.
$$
By Proposition 1 and (\ref{st2}), the inversion
$$
\tilde{S} = S^{-1}
$$ of the one-parameter  family of minimal surfaces
$$
S = \int \left(\begin{array}{cc} 0 & -1 \\ 0 & -i f^{\prime} \end{array}\right) dz +
\left(\begin{array}{cc} i \bar{f}^\prime & 0 \\ 1 & 0\end{array}\right) d\bar{z} +
+ \left(\begin{array}{cc} \bar{f}^{\prime\prime} & 0 \\ 0 & f^{\prime\prime} \end{array}\right)dt
$$
determines a DSII deformation of surfaces in $\R^4$. The matrix (\ref{kmatrix}) for these data takes the form
$$
K = \Psi S^{-1} \Gamma \Phi^\top \Gamma^{-1} =
\frac{1}{|z|^2 + |f|^2}
\left(\begin{array}{cc} \bar{z}\bar{f}^\prime - \bar{f} &
 -i\bar{z}-if^\prime \bar{f} \\
  -iz-i\bar{f}^\prime f & zf^\prime - f \end{array}\right).
$$
By Theorem 1, this transformation of surfaces induces the Moutard transformation
solutions of the trivial solution $U=V=0$ of the DSII equation to the solution
\begin{equation}
\label{solution}
U = \frac{i(zf^\prime - f)}{|z|^2 + |f|^2}, \ \
V = 2ia_z, \ \
a = -\frac{i(\bar{z} + f^\prime) \bar{f}}{|z|^2 + |f|^2}.
\end{equation}
Theorem 2 is proved.

\subsection{Examples}
\label{examples}

Let us consider the simplest examples of solutions of type (\ref{solution}).
They correspond to cases when $\varphi_1$ is a polynomial in $z$ of degree $n\leq3$.
Below by $c$ we denote  a constant $c$ which may take arbitrary complex values and by $r$ we
denote $|z|, z \in \C$.

1) $n=0$:
$$
\varphi_1 = 1, \ \ f = z+c,
$$
$$
U = -\frac{ic}{|z|^2 + |z+c|^2}, \ \ V = -\frac{2(2\bar{z}+c)^2}{(|z|^2+|z+c|^2)^2}.
$$
This solution is stationary  and for $c \neq 0$ it is nonsingular.

2) $n=1$:
$$
\varphi_1 = 2z, \ \ f = z^2 + 2it + c,
$$
$$
U = \frac{i(z^2-2it-c)}{|z|^2 + |z^2 + 2it+c|^2}, \
$$
$$
V =
\frac{4(\bar{z}^2-2it + \bar{c})}{|z|^2 + |z^2+2it+c|^2}-
\frac{2(2z(\bar{z}^2 -2it +\bar{c})+\bar{z})^2}{(|z|^2 + |z^2+2it+c|^2)^2}.
$$
The function $|U|$ decays as $O\left(\frac{1}{r^2}\right)$ as $r \to \infty$.
If $c$ is not purely imaginary, then the solution is always smooth.
It $c = it\tau, \tau \in \R$, then at $t = -\frac{\tau}{2}$ the function $U$ has a singularity at $z=0$ of the type
$$
U \sim i e^{2i \phi} \ \ \ \mbox{as $r  \to 0$}, \ \ \mbox{where $z = r e^{i\phi}$}.
$$
Remark that $U \in L_2(\R^2)$ for all $t$ and $c$.
Since a small variation of $c$ removes singularities, they are not stable.

For brevity, for the next cases we skip explicit formulas for $V$.

3) $n=2$
$$
\varphi_1(z,t) = 3z^2 + 6it, \ \ \ f = z^3 + 6itz +c,
$$
$$
U = \frac{i(2 z^3 -c)}{|z|^2 + |z^3 + 6itz + c|^2}.
$$
If $c \neq 0$, then the solution is always smooth.
If $c=0$, then it has a singularity
$$
U \sim \frac{2i}{1+36t^2} r e^{i3\phi}
$$
at $r=0$ for all $t$.

4) $n=3$:
$$
\varphi_1 = 4z^3 +24itz, \ \ f=z^4+12itz^2 - 12t^2 +c,
$$
$$
U = \frac{i(3z^4 + 12 itz^2 + 12t^2 -c)}{|z|^2 + |z^4 +12itz^2 - 12t^2 +c|^2}.
$$
This solution becomes singular for $c = 12t^2$ which is possible if and only if
$c$ is real-valued an positive. In this case it has singularities $U \sim -12t e^{2i\phi}$ at $z=0$ for $t=\pm \sqrt{c/12}$.

\section{Remarks}
\label{remarks}

1) Ozawa constructed a blowing-up solution to the DSII equation in the form (\ref{ozeq}) as follows: he took
$$
U(X,Y,0) = \frac{e^{-ib(4a)^{-1}(X^2 - Y^2)}}{a(1+ ((X/a)^2+(Y/a)^2)/2)}
$$
as the initial data for $t=0$ and showed that for constants $a$ and $b$
such that $ab<0$ we have
$$
\| U\|^2 \to 2 \pi\cdot \delta \ \mbox{as $t \to T = -a/b$}
$$
in ${\cal S}^\prime$ where $\|U\|^2 = \int_{\R^2}|U|^2\, dx\,dy$ is the squared $L_2$-norm of the solution and $\delta$ is the Dirac distribution centered at the origin. We remark that $\|U\|^2 = 2\pi$ and the solution extends for $T > -a/b$ and gains regularity. A survey of the blow-up problem for DSII is given in \cite[\S 5]{KS}.

2) The Willmore functional $\|U\|^2$ is the first integral of the system.
For (\ref{s1}) it is always equal to $2\pi$ except fot the time $T_{\mathrm{sing}}$
when solution becomes singular. For $t= T_{\mathrm{sing}}$ it is equal to
$\pi$. Analogously for (\ref{s2}) the Willmore functional is equal to $4\pi$ for $t$
such that $U$ is nonsingular and is equal to $3\pi$ for $t=T_{\mathrm{sing}}$.

The analogous effect was established for the solutions to the mNV equation constructed in \cite{T152}. The multiplicity of the value of the functional to $\pi$
in both cases is explained by that the surfaces $\widetilde{S}$ are immersed Willmore spheres (with singularities for singular moments of time).

3) A Moutard type transformation for the DSII equation was earlier constructed in  \cite{LSY}. However, it is constructed from only one spinor $\Psi$ and therefore has no geometric interpretation in terms of surface theory. The solutions constructed in \cite {LSY} have other interesting properties. Note that in \cite{MS} as in \cite{LSY} a Moutard transformation is called a (binary) Darboux transformation.

Note that the original Moutard transformation for the two-dimensional Schrodinger operator was used to construct the potentials and solutions of the Novikov--Veselov
equation with analytical properties, for example, in \cite{TT07, TT} and subsequent works.

For similar purposes, Moutard-type transformations of two-dimensional Dirac operators, except for \ cite {T152} and the recent article, have been successfully applied in \cite{GN1,GN2,GN3}.

\vskip7mm

\end{document}